\documentstyle[11pt,amstex]{article}                                %%
\def\AFOUR{%
\setlength{\textheight}{9.0in}%
\setlength{\textwidth}{5.75in}%
\setlength{\topmargin}{-0.375in}%
\hoffset=-.5in%
\renewcommand{\baselinestretch}{1.17}%
\setlength{\parskip}{6pt plus 2pt}%
}

\AFOUR                                           %%%   ***

\makeatletter
\def\section{\@startsection {section}{1}{\z@}{-3.5ex plus -1ex minus
 -.2ex}{2.3ex plus .2ex}{\large\bf}}
\def\subsection{\@startsection{subsection}{2}{\z@}{-3.25ex plus -1ex minus
 -.2ex}{1.5ex plus .2ex}{\normalsize\bf}}
\makeatother

\makeatletter
\@addtoreset{equation}{section}

\makeatother

\newcommand{\nc}{\newcommand}
\newcommand{\rnc}{\renewcommand}

\nc{\bea}{\begin{eqnarray}}
\nc{\eea}{\end{eqnarray}}
\nc{\be}{\bea}
\nc{\ee}{\eea}

\rnc{\a}{\alpha}
\nc{\ab}{\bar{\a}}
\nc{\ap}{\a^{+}}
\nc{\abm}{\ab^{-}}
\rnc{\b}{\beta}
\nc{\bb}{\bar{\b}}
\nc{\bbp}{\bb_{\zb}^{+}}
\nc{\bm}{\b_{z}^{-}}
\nc{\oa}{\overline{\a}}
\nc{\ob}{\overline{\b}}
\rnc{\gg}{\gamma}
\rnc{\d}{\delta}
\nc{\f}{\phi}
\nc{\fb}{\bar{\phi}}
\nc{\vf}{\varphi}
\nc{\p}{\psi}

\rnc{\c}{\chi}
\nc{\la}{\lambda}
\nc{\m}{\mu}
\nc{\n}{\nu}
\rnc{\o}{\omega}
\nc{\Om}{\Omega}
\rnc{\t}{\theta}
\nc{\eps}{\epsilon}
\rnc{\S}{\Sigma}
\nc{\F}{\Phi}

\nc{\trac}[2]{{\textstyle\frac{#1}{#2}}}

\nc{\ex}[1]{\mbox{e}^{\,\textstyle#1}}

\nc{\mat}[4]{\left(\begin{array}{cc}#1&#2\\#3&#4\end{array}\right)}

\nc{\som}[9]{\left(\begin{array}{ccc}#1&#2&#3\\#4&#5&#6\\#7&#8&#9%
\end{array}\right)}

\nc{\tr}{\mathop{\mbox{tr}}\nolimits}
\nc{\ad}{\mathop{\mbox{ad}}\nolimits}
\nc{\Tr}{\mathop{\mbox{Tr}}\nolimits}
\nc{\Det}{\mathop{\mbox{Det}}\nolimits}
\nc{\rk}{\mathop{\mbox{rk}}\nolimits}
\nc{\ra}{\rightarrow}
\nc{\Ra}{\Rightarrow}
\nc{\LRa}{\Leftrightarrow}
\nc{\ot}{\otimes}
\rnc{\ss}{\subset}
\nc{\nul}{\noindent\underline}
\nc{\non}{\nonumber\\}

\nc{\subs}[1]{{\vspace*{0.5cm}}%
{\noindent\underline{#1}}{\addcontentsline{toc}{subsection}{#1}}%
{\vspace*{0.3cm}}}
\nc{\zb}{\bar{z}}
\rnc{\lg}{\frak{g}}
\nc{\lt}{\frak{t}}
\nc{\lk}{\frak{k}}
\nc{\lh}{\frak{h}}
\nc{\pik}{\Pi_{\lk}}
\nc{\pip}{\Pi_{+}}
\nc{\pim}{\Pi_{-}}
\nc{\pih}{\Pi_{\lh}}
\nc{\jz}{J_{z}}
\nc{\jzh}{\jz^{\lh}}
\nc{\jzp}{\jz^{+}}
\nc{\jzm}{\jz^{-}}
\nc{\del}{\partial}
\nc{\dz}{\del_{z}}
\nc{\dzb}{\del_{\bar{z}}}
\nc{\az}{A_{z}}
\nc{\azb}{A_{\bar{z}}}
\nc{\g}{g^{-1}}
\nc{\dw}{\Delta_{W}}
\nc{\Ad}{{\mbox{Ad}}}
\nc{\ks}{Ka\-za\-ma-\-Su\-zu\-ki}
\nc{\KS}{\ks}
\nc{\ksm}{\ks\ model}
\rnc{\AA}{{\Bbb A}}
\nc{\BB}{{\Bbb B}}
\nc{\CC}{{\Bbb C}}
\nc{\PP}{{\Bbb P}}
\nc{\cpm}{\CC\PP(m)}
\nc{\cpn}{\CC\PP(n)}
\nc{\cp}[1]{\CC\PP(#1)}
\nc{\gmn}{G(m,m+n)}
\nc{\gmnk}{\gmn_{k}}
\nc{\cO}{{\cal O}}
\nc{\bcO}{\bar{\cO}}
\nc{\bO}{\bar{O}}
\nc{\oQ}{\overline{Q}}
\begin{document}
\global\parskip=4pt

%%%%%%%%% title page %%%%%%%%%%%%%%%%%%%%%%%%%%%%%%%%%%%%%%%%
\makeatletter
\begin{titlepage}
\begin{flushright}
IC/98/192
\end{flushright}
\begin{center}
\vskip .5in
{\LARGE\bf On the Generalized Casson Invariant}
\vskip 0.4in
{\bf George Thompson}\footnote{email: thompson@ictp.trieste.it}
\vskip .1in
ICTP \\
P.O. Box 586 \\
34100 Trieste \\
Italy\\

\end{center}
\vskip .4in
\begin{abstract}
The path integral generalization of the Casson invariant as developed
by Rozansky and Witten is investigated. The path integral for various
three manifolds is explicitly evaluated. A new class of topological
observables is introduced that may allow for more effective
invariants. Finally it is shown how the dimensional reduction of these
theories corresponds to a generalization of the topological {\bf B}
sigma model. 
\end{abstract}

\end{titlepage}
\makeatother
%%%%%%%%% end of title page %%%%%%%%%%%%%%%%%%%%%%%%%%%%%%

\begin{small}
\tableofcontents
\end{small}

\setcounter{footnote}{0}

\section{Introduction}
The Casson invariant, a three manifold invariant, has been with us
since 1985. Originally it was defined, by Casson, for ${\Bbb Z}$HS's
(integral homology spheres) \cite{casson}. In this situation, Taubes
showed that
the Casson invariant can be viewed as the Euler characteristic of the
Floer homology of flat $SU(2)$ connections on the integral homology sphere
\cite{taubes}. There is a path integral representation of the
invariant, due to Witten, which formally gives back the construction
of Taubes \cite{wcas}. Atiyah and Jeffrey interpreted this result as
being a definition of a `regularised' Euler characteristic of the (infinite
dimensional) space of $SU(2)$ connections on ${\Bbb Z}$HS's. An
alternative interpretation of the path integral was given in
\cite{btcas}. There it was argued that the Casson invariant is the
Euler Characteristic of the moduli space of flat connections (when the
space is disjoint it is the Euler characteristic of each component
sumed with signs given by spectral flow)\footnote{A description of
this construction which is perhaps more accessible to mathematicians can be
found in a forthcoming book \cite{mm}. There one will also find a
discussion on the three dimensional analogue of the Seiberg-Witten
invariant and its relationship to the Casson invariant.}. This
interpretation is in keeping with the work of Taubes.

The invariant was generalised to rational
homology spheres by Walker in \cite{walk}. After much activity, Lescop
gave a surgery formula for the invariant so that it could be extended
to all three manifolds, see \cite{les} and references therein. These
generalisations are for the $SU(2)$ invariant. There is now an
$SU(3)$ extension of the Casson invariant for ${\Bbb Z}$HS's due to
Boden and Herald \cite{bh}. Each generalisation is confronted by
various analytical problems which had to be overcome. The path
integral version makes sense, as it stands, for any three manifold and
for any gauge group and so, in principal, offers a handle on the
Casson invariant that goes beyond what is mathematically accesable at
present. Unfortunately the path integral in question proved very
difficult to evaluate. 

The situation changed dramatically as a consequence of the solution of
the $N=2$ super Yang-Mills theory in four dimensions \cite{sw4d}. This
work had profound consequences for the study of four manifold
invariants. On passing to three dimensions\footnote{In section
\ref{red} I explain, rather broadly, how one passes from a higher dimensional
manifold to a lower dimensional one.} Seiberg and Witten
\cite{sw3d} gave a solution to the $N=4$ super Yang-Mills theory with
gauge group $SU(2)$ in the coulomb branch. The moduli space of the
theory was conjectured to be the Atiyah-Hitchin two monopole moduli
space. To pass to the path integral representation of the Casson
invariant one starts with the physical theory and one twists it (I
describe this in the body of the paper). Since the topological theory
ought not to depend on which scale we are looking at, twisting the
full theory or twisting the low energy effective theory should yield
the same invariant. The Casson invariant is therefore given by a
particular path integral part of whose data includes the integration
over the space of maps from the three manifold to the Atiyah-Hitchin
space $X_{{\mathrm AH}}$. 

The Rozansky-Witten invariant $Z^{RW}_{X}[M]$ is then one
manifestation of the path integral invariant \cite{rw}, the manifestation in
which one has a supersymmetric sigma model of maps from $M$ to some
hyper-K\"{a}hler manold $X$. This corresponds to the Casson invariant
of $M$ when $X= X_{{\mathrm AH}}$. Rozansky and Witten establish that
in this case one does indeed reproduce all the known general results
of the Casson invariant including the surgery formula of Lescop
\cite{les}. One can read this sucess in the opposite direction, namely
that this confirms the conjecture of Seiberg and Witten that for the
physical theory the coulomb branch moduli space is $X_{{\mathrm AH}}$.

Some immediate
consequences of \cite{rw} are that for $b_{1}(M)>3$ the invariant
vanishes and for
$1\leq b_{1}(M) \leq 3$ the invariant is related to classical
invariants of the three manifold. So while one has a generalization of
the Casson invariant, this generalization does not, at first sight,
help to provide non-trivial invariants for three manifolds with $b_{1}(M)
\neq 0$. One of the aims of this paper is to introduce observables
which may correspond to non-trivial invariants for any $b_{1}(M)$.

In section \ref{rwm} I give a quick review of the work of Rozansky and
Witten, though the discussion presented is slightly different from the
one in \cite{rw}. This is followed, in section \ref{calcs}, with a
non-perturbative evaluation of the path integral for manifolds of the
form $\Sigma\times S^{1}$. One obtains, in a straightforward way, the
invariant $Z^{RW}_{X}[\Sigma\times S^{1}]$ in terms of invariants of
$X$. One advantage of the approach adopted here is that one does not
need to know that the Hilbert space of states is finite
dimensional. The following section proceeds to slightly more
complicated manifolds-mapping tori. This is followed by a discussion
of the types of theories that one obtains on dimensional reduction. One
finds B-type topological sigma models. Kapranov \cite{kap} and
Kontsevich \cite{kon} have shown that the manifold $X$ need not be
hyper-K\"{a}hler it is enough that it be holomorphic
symplectic. Indeed the Kapranov-Kontsevich
theory reduced yields a slight generalisation of the topological B-models.

There are a number of things that are missing. One has to do with the
relationship of the Rozansky-Witten invariants to the known universal
invariant the so called LMO invariant \cite{lmo}. Rozansky and Witten
establish that their theory provides a weight system, but this is just short
of establishing that their invariants arise from the LMO
invariant. The missing part is given in
\cite{ht}, where a rather more general evaluation of the path
integrals involved is also to be found. The relationship to the
$SU(3)$ invariant \cite{bh} is also adressed there.

Another glaring gap is the relationship between this work and
Donaldson theory on four manifolds of the form $M\times
S^{1}$, the Seiberg-Witten invariant and the {\bf u}-plane. Something
which will be filled in elsewhere as it is one of the main motivations
for this work.

\section{The Rozansky-Witten Model}\label{rwm}
The starting point is that at low energies the path integral that
corresponds to the Casson invariant becomes a topological
supersymmetric sigma model. The supersymmetry is such that the target
space of the sigma model should be hyper-K\"{a}hler. To write down
such a topological field theory in three dimensions Rozansky and
Witten twist a model that comes from the reduction of a supersymmetric
sigma model in six dimensions. The way this works is reviewed
presently. For the moment all we need are some facts about
hyper-K\"{a}hler manifolds and their complexified tangent bundles. 

A real manifold of dimension $2m$ has holonomy group ${\mathrm
SO}(2m)$ (or some
subgroup thereof). If that manifold admits a complex structure then
the holonomy group can be reduced to ${\mathrm U}(m) \subset
{\mathrm SO}(2m)$. Furthermore if the manifold admits a Ricci flat metric the
holonomy group is ${\mathrm SU}(n)$. If $X$ is a hyper-K\"{a}hler manifold
($\dim_{{\Bbb R}}X=4n$) then
there is a Riemannian metric such that the Levi-Cevita connection lies
in an ${\mathrm Sp}(n)$ subgroup of ${\mathrm SO}(4n)$ (so that
${\mathrm Sp}(n) \subseteq {\mathrm SU}(2n) \subset {\mathrm U}(2n) \subset
{\mathrm SO}(4n)$). The
complexified tangent bundle decomposes as
\be
TX_{{\Bbb C}} = TX \otimes_{{\Bbb R}} {\Bbb C} = V \otimes S ,
\label{decomp} 
\ee
where $V$ is a rank $2n$ complex vector bundle with structure group
${\mathrm Sp}(n)$ and $S$ is a trivial rank 2 complex vector bundle
with structure group ${\mathrm Sp}(1)$. The Levi-Cevita connection is
a connection in $V$ and the trivial connection in $S$. ${\mathrm
Sp}(1)$ labels are A, B, $\dots$ and there is an invariant
antisymmetric tensor $\eps_{AB}$ with inverse $\eps^{AB}$,
\be
\eps^{AC}\eps_{CB} = \d^{A}_{B} .
\ee
${\mathrm Sp}(n)$ labels are I, J, $\dots$ and there is also an
invariant antisymmetric tensor $\eps_{IJ}$ with inverse $\eps^{IJ}$,
\be
\eps^{IK}\eps_{KJ} = \d^{I}_{J}.
\ee 

Local coordinates on $X$ will be denoted $\f^{i}$ and the metric is
$g_{ij}$. The fact that the tangent bundle decomposes as in
(\ref{decomp}) means that there exist covariantly constant tensors
$\gg^{AI}_{i}$ and $\gg_{AI}^{i}$ that describe the maps from $V
\otimes S$ to $TX_{{\Bbb C}}$ and vice versa. Using these tensors one
may express the Riemann curvature tensor as
\be
R_{ijkl} = - \gg^{AI}_{i} \gg^{BJ}_{j} \gg^{CK}_{k} \gg^{DL}_{l}
\eps_{AB} \eps_{CD} \Omega_{IJKL} ,
\ee
where $\Omega_{IJKL}$ is completely symmetric in the indices. A useful
relationship is
\be
\eps_{IJ}\gg^{AI}_{i}\gg^{BJ}_{j} = \frac{1}{2}g_{ij}\eps^{AB} .
\ee

Fix on a complex structure so that $\f^{I}$ are holomorphic
coordinates on $X$ with respect to this complex structure. Then we may
take
\be
\gg_{A J}^{I} = \d_{A1} \d^{I}_{J}, \;\;\;\; \gg_{A I}^{\overline{I}}
= \d_{A 2} g^{\overline{I}J}\eps_{JI}.
\ee
Notice that this means that we have essentially chosen
$V_{{\Bbb C}}=T^{(1,0)}X$. $S$ is a trivial rank 2 bundle so that on
picking a trivialisation one has $S = {\Bbb C} \oplus {\Bbb C}$. Now
this means that $T_{{\Bbb C}}X = V \otimes S = V \otimes \left({\Bbb C}
\oplus {\Bbb C} \right) \equiv T^{(1,0)}X \oplus T^{(0,1)}X$.

With all the preliminaries out of the way we can write down the action.
The action is
\be
S = c_{1}\int_{M} L_{1} \sqrt{h} d^{3}x\, + \, c_{2}\int_{M} L_{2}
\label{rwact} 
\ee
where
\bea
L_{1} & = & \frac{1}{2}g_{ij}\partial_{\mu}\f^{i}\partial^{\mu}\f^{j}
+ \eps_{IJ} \chi^{I}_{\mu}D^{\mu} \eta^{J} \label{lag1}\\
L_{2} & = & \frac{1}{2}\left( \eps_{IJ} \chi^{I} D \chi^{J} +
\frac{1}{3} \Omega_{IJKL} \chi^{I}\chi^{J}\chi^{K} \eta^{L}
\right). \label{lag2} 
\eea
The covariant derivative is
\be
D_{\mu \, J}^{I} = \partial_{\mu}\d^{I}_{J} +
(\partial_{\mu}\f^{i})\Gamma^{I}_{i J} .
\ee

Both of the Lagrangians $L_{1}$ and $L_{2}$ are invariant under two
independent BRST supersymmetries. These supersymmetries are defined without
the need of picking a prefered complex structure on $X$. We will not
need this level of generality here, so pick a complex structure on $X$
so that the $\f^{I}$ are local holomorphic coordinates with respect to
this complex structure and let one of the BRST operators,
$\overline{Q}$ act by
\be
\begin{array}{ll}
\overline{Q}\f^{I} = 0, & \overline{Q}\f^{\overline{I}} =
g^{\overline{I}J} \eps_{JK} \eta^{K}, \\
\overline{Q}\eta^{I}= 0, & \overline{Q}\chi^{I} = -d\f^{I} .
\end{array} \label{brst}
\ee

Since $L_{1}$ is BRST exact we may vary the coefficient $c_{1}$ with
impunity. In particular one can take it, as the authors of \cite{rw}
do, to equal $c_{2}$. There is an alternative choice which, from a
different point of view, is more natural. If one starts with the four
dimensional u-plane theory of Moore and Wiiten \cite{mw} on a four
manifold of the form $M\times S^{1}$ where the circle has radius $R$
(tending to zero), then on reduction one would find $c_{1}= 1/R$ and
$c_{2}=1$. As far as the partition function is concerned it is
irrelevant as to which choice we make for the coefficients. However,
the choices made affect the normalization of observables as will be
seen below. For definiteness and to keep contact with \cite{rw} in
most of what follows I will take, $c_{1}=c_{2}=c$.

\subsection{The Measure}

To completely specify the theory one needs to fix on some measure for
the path integral. The theory under consideration does not have
equality between Grassman
odd and Grassman even fields, so it would seem that there is no
canonical choice of measure for the path integral. However, there are
a number of ways to see that the coupling constant dependence of
the measure should be
\be
D\left(\f^{i}\sqrt{c_{1}}\right)
D\left(\eta^{I}\frac{c_{1}}{\sqrt{c_{2}}} \right)
D\left(\chi^{I} \sqrt{c_{2}} \right) . \label{meas}
\ee
Firstly, this is the standard measure that one would choose given the
kinetic terms in (\ref{rwact}). Secondly, in the topological gauge theory
that corresponds to the Casson invariant there is a precise pairing
between Grassmann even and Grassmann odd fields which means that the
measure in that theory is such that it does not have coupling constant
dependence. Some of the fields that appear there are multiplier fields
and these can be integrated out (algebraically eliminated). The net
effect of integrating out the multiplier fields is to put the coupling
constant into the measure of the remaining fields precisely as
indicated in (\ref{meas}) (for the gauge theory). Also, thinking of
${\mathrm L}_{2}$ as a Chern-Simons type topological Lagrangian
density and ${\mathrm L}_{1}$ as a gauge fixing Lagrangian one would
be led to the measure (\ref{meas}), which guarantees that at one loop
there is no coupling constant dependence.

The choice of measure is important since it will be crucial for us to
follow the coupling constant dependence as we take various
limits. From now on set $c_{1}=c_{2} = c$.

Apart from factors of $c$ there are questions concerning the $\eps$
dependence of the measure. In \cite{rw} a detailed specification of
this dependence is given. Also in the text some factors of $2\pi$ make
an appearance that is not commented on, for example in (\ref{R}), these
factors are obtained on scaling the relevant zero modes by
$\sqrt{2\pi}$ and are consistent with the normalization of the path
integral given in \cite{rw}.

\subsection{Reduction of 6-Dimensional Theories and Twisting}
The way that Rozansky and Witten derived the action (\ref{rwact}) was
to begin with a supersymmetric sigma model in six dimensions, to
reduce to three and then to twist. The model in question has a map
$\Phi : {\Bbb R}\rightarrow  X$ and chiral fermions $\psi$ taking
values in $\Phi^{*}(V)$. The $SO(6)$ (${\mathrm spin}(6)$) Lorentz
symmetry for the bosons (fermions) in six dimensions becomes an
$SO(3)_{E} \times SO(3)_{N}$ ($(SU(2)_{E} \times SU(2)_{N})/{\Bbb
Z}_{2}$) symmetry on dimensional reduction\footnote{In the following
discussion I will not be careful about wether the reduction
includes a time like direction or not. Subtleties that arise from this
have been adressed in \cite{btesym}.} to three dimensions, where
$SO(3)_{E}$ is the Lorentz group on ${\Bbb R}^{3}$. The supersymmetry
charges transform as two copies of $({\mathbf 2},{\mathbf 2})$ under
$(SU(2)_{E} \times SU(2)_{N})$. Such a supersymmetry cannot be placed
on an arbitrary three manifold. To proceed one
`twists'. Twisting in
this case means taking the diagonal subgroup $SU(2)_{E'}$ of $(SU(2)_{E} \times
SU(2)_{N})$ to be the new Lorentz group. Under $SU(2)_{E'}$ 
each copy of the supersymmetry charges transforms as ${\mathbf 1}
\oplus {\mathbf 3}$. The tensorial properties of the fields can be
determined. Note that under $SO(3)_{E} \times SO(3)_{N}$ the
field $\Phi$ transforms as $({\mathbf 1}, {\mathbf 1})$  under
$SU(2)_{E'}$ it is therefore a ${\mathbf 1}$, so that it remains a
zero form. The spinor $\psi$ transformed as $({\mathbf 2},{\mathbf 2})$
under $(SU(2)_{E} \times SU(2)_{N})/{\Bbb Z}_{2}$ so
that under the diagonal it transforms as ${\mathbf 1}
\oplus {\mathbf 3}$, splitting into the zero form $\eta$ and the one
form $\chi$.

The gauge theory that produces the Casson invariant is a twisted $N=4$
supersymmetric theory. This can also be obtained by reducing a theory
from six dimensions down to three. The theory in question is $N=1$
super Yang-Mills theory, which enjoys, in six dimensions, an extra
`accidental' $SU(2)_{R}$ symmetry. The field content in six dimensions
is a connection $A_{M}dx^{M}$ and a chiral spinor $\psi$ (a section of
${\mathcal S}_{+} \otimes {\mathrm adP}$, where ${\mathcal S}_{+}$ is
the positive spin bundle). The
supersymmetry charge, on reduction to
three dimensions, transforms as a $({\mathbf 2},{\mathbf 2},{\mathbf 2})$
of $SU(2)_{E} \times SU(2)_{N} \times SU(2)_{R}$. As one can imagine
there are many possible twisted theories, however the one of interest
does not come by taking the diagonal of $SU(2)_{E} \times SU(2)_{N}$
as for the sigma model, rather, one sets  $SU(2)_{E'}$ to be the
diagonal of $SU(2)_{E} \times SU(2)_{R}$. The supersymmetry charge remains a
doublet of $SU(2)_{N}$, and transforms as a ${\mathbf 1} \oplus
{\mathbf 3}$ of $SU(2)_{E'}$. On reduction the connection transforms
as $({\mathbf 3},{\mathbf 1},{\mathbf 1}) \oplus ({\mathbf 1},{\mathbf
3},{\mathbf 1})$ while the spinor transforms as $({\mathbf 2},{\mathbf
2},{\mathbf 2})$ of $SU(2)_{E} \times SU(2)_{N} \times SU(2)_{R}$. On
twisting, the connection transforms as $({\mathbf 3},{\mathbf 1}) \oplus
({\mathbf 1},{\mathbf 3})$ and the spinor as a $({\mathbf 1},{\mathbf 2})
\oplus ({\mathbf 3},{\mathbf 2})$ under $SU(2)_{E'} \times SU(2)_{N}$.

It is curious that the two different path integral realizations of the
Casson invariant arise on taking different twists of theories which
start of life in six dimensions.

\subsection{The Perturbative Limit} \label{pert}
The fact that the theory does not depend on the constant $c$, means
that we will be able to `localise' the path integral onto the set of
constant maps. Let,
\be
c = \frac{1}{R} \label{coef},
\ee
with the aim of, shortly, taking the $R \rightarrow 0$ limit. Write
the fields $\f^{i}$ and $\eta^{I}$ in an orthogonal decomposition as
`fluctuating' parts with subscript $q$ (for quantum) plus harmonic
components which
have the subscript $0$ 
\bea
\f^{i} &= & \f^{i}_{0}+ \f^{i}_{q} \\
\eta^{I} &= & \eta^{I}_{0}+  \eta^{I}_{q}  ,
\eea
($\chi^{I}$ can still be harmonic). In order to take the $R
\rightarrow 0$ limit we scale the fluctuating
fields
\bea
\f^{i} &\rightarrow &  \f^{i}_{0} +  \sqrt{R} \f^{i}_{q} , \nonumber \\ 
\eta^{I} &\rightarrow &  \frac{1}{\sqrt{R}} \eta^{I}_{0} + \sqrt{R}
\eta^{I}_{q}, \nonumber \\
\chi^{I} &\rightarrow &\sqrt{R} \chi^{I} . \label{scales}
\eea
Notice that this scaling does not have unit Jacobian, rather the
effect is to change the measure (\ref{meas}) to 
\be
D\left(\f^{i}\right) D\left(\eta^{I}\right)
D\left(\chi^{I}\right) .
\ee

On taking the limit the resulting
Lagrangians are\footnote{One also needs to shift 
\be
\eta^{I} \rightarrow
\eta^{I} - \frac{1}{R}\f^{i}\Gamma_{iJ}^{I}(\f_{0})\eta^{J}_{0}.
\ee}

\bea
L_{1} &\rightarrow &
\frac{1}{2}g_{ij}(\f_{0})\partial_{\mu}\f^{i}_{q}\partial^{\mu}\f^{j}_{q} + 
\eps_{IJ}(\f_{0}) \chi^{I}_{\mu}\partial^{\mu} \eta^{J}_{q} \nonumber \\
& & \;\;\;\;-\frac{1}{2}
\gamma_{i}^{AK}\gamma_{j}^{BL}
\eps_{AB}\Omega_{IJKL}(\f_{0})\chi_{\mu}^{I}\eta^{J}_{0}
\f^{j}_{q}\partial^{\mu} \f^{i}_{q} \\
L_{2} & \rightarrow & \frac{1}{2}\left( \eps_{IJ}(\f_{0}) \chi^{I} d \chi^{J} +
\frac{1}{3} \Omega_{IJKL}(\f_{0}) \chi^{I}\chi^{J}\chi^{K} \eta^{L}_{0} \right)
\eea
and these agree with the Lagrangians found in \cite{rw} which
correspond to what they termed minimal Feynman diagrams\footnote{With
$c_{1}=1/R$ and $c_{2}=1$ the required scalings to arrive at the
minimal Feynman vertices are $\f^{i}
\rightarrow   \f^{i}_{0} +  \sqrt{R} \f^{i}_{q} $ and $\eta^{I}
\rightarrow  \eta^{I}_{0} + R \eta^{I}_{q}$. \label{sc2}}.

\subsection{Observables}

Rozansky and Witten introduced two types of observables. The first is
made up of only the $\eta$ field and certain classes on $X$. There is
an isomorphism between the spaces $\Omega^{(k,0)}(X)$ and
$\Omega^{(0,k)}(X)$ given by the tensor
\be
S^{I}_{\overline{J}} = \eps^{IJ}g_{J \overline{J}} .
\ee
Let $\o$ be a $k$-form which is $\partial$ closed as a $(k,0)$-form
and which is $\overline{\partial}$ closed as a $(0,k)$-form. From this
and the similarity of the BRST operators to the Dolbeault operators
one concludes that
\be
{\mathcal O}(\o) = \o_{I_{1}, \dots , I_{k}}(\f) \, \eta^{I_{1}}\dots
\eta^{I_{k}} ,
\ee
is BRST closed and hence a good observable in the theory. From the
scalings that we performed in the previous section we learn that the
these observables scale with an $R$ dependence, as
\be
{\mathcal O}(\o) \mapsto \left(\frac{1}{\sqrt{R}} \right)^{k} \,
\o_{I_{1},  \dots , I_{k}}(\f_{0}) \,
\eta^{I_{1}}_{0}\dots \eta^{I_{k}}_{0} + \dots , 
\ee
where the ellipses indicate lower order terms. For example, when $n=1$
and $k=2$, the observable scales as
\be
{\mathcal O}(\o) = \frac{1}{R}\o_{I_{1}, I_{2}}(\f_{0}) \, \eta^{I_{1}}_{0}\,
\eta^{I_{2}}_{0} + \dots .
\ee
These observables play a role in determining, from the path
integral point of view, the behaviour of the Casson invariant under
the operation of taking a connected sum \cite{rw}. The normalisation of the
observable, as I have given it here, therefore seems to be
incorrect. Rather, one should perhaps define it from the outset to be
\be
{\mathcal O}(\o) = \left(\sqrt{R}\right)^{k}\o_{I_{1}, \dots ,
I_{k}}(\f) \,  \eta^{I_{1}}\dots
\eta^{I_{k}} ,
\ee
though I have no convincing argument for this from the three
dimensional viewpoint at present. On the other hand if we had begun
with $c_{1}=1/R$ and $c_{2}=1$ then after the scalings indicated in
footnote \ref{sc2} we would have found the more appealing result
\be
{\mathcal O}(\o) \rightarrow \o_{I_{1}, \dots ,
I_{k}}(\f_{0}) \,  \eta^{I_{1}}_{0}\dots
\eta^{I_{k}}_{0} .
\ee

Let ${\mathcal K}$ be a knot and ${\mathcal R}$ a representation of
Sp$(n)$. The second set of observables are constructed from the gauge field,
\be
A_{IJ} = d\f^{i} \eps_{IK}\Gamma^{K}_{iJ} + \Omega_{IJKL}\chi^{K}\eta^{L} .
\ee
Under a BRST transformation $A$ transforms as a gauge field ought to,
so that Wilson loops,
\be
{\mathcal W}_{{\mathcal R}}({\mathcal K}) = {\mathrm Tr}_{{\mathcal R}} \, \ex
{ \oint_{{\mathcal K}} A}
\ee
are BRST invariant and metric independent. Under the scalings of the
last section (also for $c_{1}=1/R$ and $c_{2}=1$) the gauge field goes
over to
\be
A_{IJ} \mapsto \Omega_{IJKL}(\f_{0})\chi^{K}\eta^{L}_{0 }. \label{conn}
\ee
This connection will make an appearance again when we come to Mapping-Tori.

\subsection{New Invariants and Beyond the $b_{1}(M)=3$ Barrier}

The problem that one faces in obtaining non-zero results past
$b_{1}(M)=3$ is that there seems to be no natural way in which to
handle the (excess) $\chi^{I}$ harmonic modes. For ${\mathrm
dim}_{\Bbb R}X = 4n$ and $b_{1}(M)=b_{1}$ there are $2nb_{1}$ $\chi$ harmonic
modes and $2n$ $\eta$ harmonic modes. The $\eta$ harmonic mode count
tells us that the total number of insertions of vertices in any given
diagram is $2n$ so that that the most one can hope to do is to soak up
$6n$ $\chi$ harmonic modes (since, as one can see from the vertices,
there are at most three $\chi$ for each $\eta$). This puts the upper limit on
$b_{1}$. Insertions of the observables do not help matters. In fact
both the point
observables ${\mathcal O}(\o)$ and the knot observables ${\mathcal
W}({\mathcal K})$ work the wrong way in that they are $\eta$ dependent.

Apart from the two types of
observables that have already been discussed there 
is a third type which is available when $b_{1}(M) \neq 0$ and which
can ``soak up'' the $\chi^{I}$ zero modes. However,
the observables that we will construct are only invariant under one of
the BRST operators. This should not be a problem in making sense of
the observables. The usual arguments to prove metric independence of
the expectation value of a product of observables only requires the
existance of one BRST operator, $\overline{Q}$, with respect to which
the observables in question are closed. If the observable is
$\overline{Q}$ exact, then under normal conditions, its  expectation
value will vanish. If the $\overline{Q}$ invariant observable has
non-zero expectation value then we may conclude that it is a
non-trivial topological observable. The r\^{o}le of the other BRST
operator is to pick out a representative of the $\overline{Q}$ class
that the observable belongs to. The problem in explicitly constructing
such a representative in the field theory is that it may have to be
non-local. Of course, at the level of cohomology, any representative of
a class is as good as any another so we do not need to impose the
extra condition that observables are also $Q$ closed. Hence, I will
take the attitude that since the observables below are
in $\overline{Q}$ cohomology and are independent of the metric on
$M_{3}$, that they constitute good topological observables.

The $\overline{Q}$ supersymmetry is such that given any closed $1$-cycle
$\gg$ 
\be
\chi_{I}(\gg) = \eps_{IJ}\oint_{\gg} \, \chi^{J} ,
\ee
is invariant. The path integral is such that essentially the
$\chi^{I}$ of interest will be harmonic on $M$ so that the $\gg$ may
as well be taken to live in ${\mathrm H}_{1}(M)$. For each $\gg$ one
may view $\chi_{J}(\gg)$ as a section of $T^{(1,0)}X$. Let $\gg_{i}$,
$i= 1, \dots, b_{1}(M)$, be a basis of $H_{1}(M)$ and set
\be
\chi_{I}(\gg_{i}) = \eps_{IJ}\oint_{\gg_{i}} \, \chi^{J} .
\ee
One also has
that
\be
\eta^{\overline{I}} = g^{\overline{I}J}\eps_{JK}\eta^{K} ,
\ee
is $\overline{Q}$ invariant and the $\eta^{\overline{I}}$ are sections
of $T^{(0,1)}X$.

Let $\la$ be a $(0,q)$ form on $X$ with values in $\wedge^{k_{1}}
T^{(1,0)}X \otimes \wedge^{k_{2}} T^{(1,0)}X \otimes \dots \otimes
\wedge^{k_{r}} T^{(1,0)}X$ where $b_{1}(M)=r$. In local coordinates
such an object can be written as
\be
\la_{\overline{I}_{1}\overline{I}_{2} \dots \overline{I}_{q} }^{
(J_{1}^{1}\dots J_{k_{1}}^{1}), \dots ,
(J_{1}^{r}\dots J_{k_{r}}^{r})}\; d\overline{z}^{\overline{I}_{1}}
 \dots
d\overline{z}^{\overline{I}_{q}} \otimes \frac{\partial}{\partial
z^{J_{1}^{1}}} \dots \frac{\partial}{\partial
z^{J_{k_{1}}^{1}}} \otimes \dots \otimes \frac{\partial}{\partial
z^{J_{1}^{r}}} \dots \frac{\partial}{\partial
z^{J_{k_{r}}^{r}}} .
\ee
In the above formula labels in a given set are understood to be
antisymmetric amongst themselves. Now given such a form one can
construct an observable
\bea
{\mathcal O}(\la)\left(q;k_{1}, k_{2}, \dots , k_{r} \right)
& = & \la_{\overline{I}_{1}\overline{I}_{2}  \dots \overline{I}_{q} }^{
(J_{1}^{1}\dots J_{k_{1}}^{1}), \dots ,
(J_{1}^{r}\dots J_{k_{r}}^{r})} 
\; \eta^{\overline{I}_{1}} \dots \eta^{\overline{I}_{q}} . \nonumber \\
&  & \;\;\;\; \chi(\gg_{1})_{J_{1}^{1}} \dots
\chi(\gg_{1})_{J_{k_{1}}^{1}}  \dots
\chi(\gg_{r})_{J_{1}^{r}}\dots \chi(\gg_{r})_{J_{k_{r}}^{r}}.
\eea 
One may well need to average over cycles in computing expectation
values of such observables to ensure that the outcome does not depend
on the choices of basis that we have made.

Perhaps the prototypical example of such an observable is that made
out of the holomorphic symplectic two form itself. One defines
\be
\eps(\gg_{1}, \gg_{2}) = \eps_{IJ}\oint_{\gg_{1}}\chi^{I}
\oint_{\gg_{2}}\chi^{J} ,
\ee
which is BRST invariant by virtue of the fact that $\eps_{IJ}$ is
holomorphic. Products of such operators are also BRST invariant and
hence good observables. Each such operator soaks up $2$ of the $\chi$ zero
modes so that if there are $m$ such operators then one finds the
`selection rule'
\be
6n+2m = 2b_{1}n
\ee
which acts as an upper bound on $b_{1}$ so that the expectation value of
the observable does not necessarily vanish. 

Let $\la = \eps^{n}$ and note that $\la \in \Omega^{(0,0)}(M,
\wedge^{2n} T^{(1,0)}X)$ so that it is of the type that we have
introduced above. One may define observables
\be
{\mathcal O}(\la)(\gg_{i}) = \left
( \eps_{IJ}\oint_{\gg_{i}}\chi^{I}\oint_{\gg_{i}}\chi^{J} \right)^{n} .
\ee

For example let $b_{1}=4$,
then we need $m=n$. In fact a nice example is $n=m=2$.

These observables may give rise to more effective invariants of three
manifolds with $b_{1}>0$ than the Rozansky-Witten invariants. As
discussed in the introduction $Z_{X}^{RW}[M]$ is, for all hyper-K\"{a}hler
manifolds $X$, essentially a classical invariant of $M$ if
$b_{1}(M)>0$. How one deals with the Rozansky-Witten invarant in
perturbation theory depends on $b_{1}(M)$ through the number of
$\chi^{I}$ harmonic modes. Insertions of operators that only involve the
$\chi^{I}$ harmonic modes means that the perturbation series is
effectively the same as that for a three manifold with smaller first
betti-number. Inserting enough operators into the path integral will
mean that the invariant will be proportional to the same type of
integrals of products of Greens functions that are typical of the
Casson invariant for a three manifold with $b_{1}=0$.

The choice of normalization of operators is important here as
well. With the normalization above and the scalings of the previous
section these observables will vanish as $R \rightarrow 0$. For the
choice $c_{1}=1/R$ and $c_{2}=1$ the scalings given in footnote \ref{sc2}
tell us to replace the fields with their harmonic modes and in the
$R \rightarrow 0$ limit the operators survive.

\subsection{Ray-Singer Torsion and Duality}
There is a small puzzle that presents itself. The twisted $N=4$ theory
will calculate the Casson invariant. One can twist the physical low
energy theory and once more one should be calculating the Casson
invariant. But now we seem to have gotten more than our moneys worth
as not only does one find the Casson invariant but at one loop the
theory also yields up the Ray-Singer torsion. How can this be?

The answer to this is that there is a difference between the
topological theory that one finds by twisting the low energy effective
physical theory and the Rozansky-Witten model. They differ by
dualising the gauge field. The path integral
for a photon and that for a (compact scalar) on ${\Bbb R}^{3}$ agree,
however, on an arbitrary three manifold the ratio of path integrals is
precisely the Ray-Singer torsion. The easiest way to see this is to
simply take the ratio. The path integral for the gauge field gives
\be
\det{\Delta_{1}}^{-1/2}.\det{\Delta_{0}},
\ee
while that for the scalar is
\be
\det{\Delta_{0}}^{-1/2}
\ee
where the subscript denotes the form degree that the Laplacian is
acting on. The required ratio is
\be
\det{\Delta_{1}}^{-1/2}.\det{\Delta_{0}}^{3/2}
\ee
which is (the inverse of) the Ray-Singer Torsion.
So on dualising the topological
gauge theory to obtain the topological sigma model one is feeding in
the one loop Ray-Singer Torsion.

\section{Calculations on $\Sigma_{g} \times S^{1}$}\label{calcs}

When the three manifold is a product of a Riemann surface $\Sigma_{g}$,
of genus $g$, and a circle one can explicitly evaluate the
Rozansky-Witten path integral. The actions on such a manifold are,
\bea
S_{1} & = & \int_{\Sigma_{g} \times S^{1}} \, \sqrt{h_{2}} \,
\frac{1}{2} g_{ij}\left
( \partial_{\mu}\f^{i}\partial^{\mu}\f^{j} +
\partial_{t}\f^{i}\partial_{t}\f^{j} \right)
+ \eps_{IJ} \left( \chi^{I}_{\mu}D^{\mu} \eta^{J} + 
\overline{\eta}^{I}D_{t} \eta^{J} \right)  \label{sig1}\\
S_{2} & = & \int_{\Sigma_{g} \times S^{1}} \, \left(\eps_{IJ}\left
( \overline{\eta}^{I} D \chi^{J} -
\frac{1}{2} \chi^{I}D_{t} \chi^{J} \right)
+ \Omega_{IJKL} \chi^{I}\chi^{J}\overline{\eta}^{K} \eta^{L} \right) ,
\label{sig2}
\eea
where $x^{\mu}$ are local co-ordinates on $\Sigma_{g}$, I have taken a
product metric
\be
h_{3} = h_{2} \oplus dt^{2} ,
\ee
and have set $\chi_{t}^{I} = \overline{\eta}^{I}$. 

In order to proceed we will
need to scale the metric of the Riemann surface down so as to obtain
an effective super quantum mechanics theory. Keep the overall coefficient
(\ref{coef}) as is but scale the metric
\be
ds^{2} = R \, h_{\mu \nu}dx^{\mu}\otimes dx^{\nu} \oplus dt^{2} ,
\ee
so that this limit is slightly different to the one we adopted for the
perturbative calculations. Of course $S_{2}$ does not feel these choices,
but $S_{1}$ certainly does. In order to be able to take the $R
\rightarrow 0$ limit one needs to scale the fields. Let a $0$
subscript denote the part of the fields which are harmonic on the
$\Sigma_{g}$ and fields with a $\perp$ subscript are orthogonal to
these, with respect to the metric on $\Sigma_{g}$. The scalings that
are to be made are the following
\bea
\f^{i} &\rightarrow & \f^{i}_{0} + \sqrt{R} \f^{i}_{\perp} \nonumber \\
\chi^{I} &\rightarrow & \sqrt{R}\left(\chi^{I}_{0} + \chi^{I}_{\perp}
\right) \nonumber \\
\eta^{I} &\rightarrow &\eta^{I}_{0}+ \sqrt{R}\eta^{I}_{\perp} \nonumber \\
\overline{\eta}^{I} &\rightarrow & \overline{\eta}^{I}_{0}+
\sqrt{R}\overline{\eta}^{I}_{\perp} . \label{sc}
\eea
After these scalings the limit $R\rightarrow 0$ can be taken. Once
more the choice of scaling is so that the measure is now
\be
D\left(\f^{i}\right) D\left(\eta^{I}\right)D\left(\overline{\eta}^{I}\right)
D\left(\chi^{I}\right) .
\ee

Let us look at each term seperately. As far as $cS_{1}$ is
concerned the bosonic field kinetic energy
terms are well behaved giving
\be
\int_{\Sigma_{g} \times S^{1}} \, \sqrt{h_{2}}\, \frac{1}{2} \, g_{ij}
(\f_{0})
\left( \partial_{\mu}\f^{i}_{\perp}\partial^{\mu}\f^{j}_{\perp} +
\partial_{t}\f^{i}_{0}\partial_{t}\f^{j}_{0} \right). \label{first}
\ee
There are also the fermionic terms
\bea
& & \frac{1}{R}\int_{\Sigma_{g} \times S^{1}} \sqrt{h_{2}}\, 
\eps_{IJ}(\f_{0})\; \chi^{I}_{\mu}D^{\mu} 
\eta^{J} \nonumber \\
& & \;\;\;\;\; \rightarrow \int_{\Sigma_{g} \times S^{1}} \, \sqrt{h_{2}}
\eps_{IJ}(\f_{0})\; \chi^{I}_{ \mu} \, \partial^{\mu}\left
( \eta^{J}_{\perp}+ \f^{i}_{\perp}\Gamma^{J}_{iK}(\f_{0}) \eta^{K}_{0}
\right) , \label{lim}
\eea
and
\be
\frac{1}{R}\int_{\Sigma_{g} \times S^{1}} \sqrt{h_{2}}\,
\eps_{IJ}\; \overline{\eta}^{I}D_{t} \eta^{J} 
%%\nonumber \\
%%& & \;\;\; 
\rightarrow 
\int_{\Sigma_{g} \times S^{1}} \sqrt{h_{2}}\,  \eps_{IJ}(\f_{0})
\overline{\eta}_{0}^{I}D_{t}(\f_{0}) \eta^{J}_{0} 
%+ \f^{j}_{\perp}
%\partial_{j} \eps_{IJ}(\f_{0}) \overline{\eta}^{I}_{\perp}
%D_{t}(\f_{0})  \eta^{J}_{0} \right. \nonumber \\
%& & \left. \; \;\;\;\; \;\;\;\;\; + \; 
%\eps_{IJ}(\f_{0}) \overline{\eta}^{I}_{\perp}
%\partial_{t}\f^{i}_{\perp} \eta^{J}_{0} +
%\eps_{IJ}(\f_{0})\overline{\eta}^{I}_{\perp}
%D_{t}(\f_{0})\eta^{J}_{\perp} \right) 
. \label{lim1}
\eea

Now turn to $cS_{2}$. We have,
\bea
\frac{1}{R}\int_{\Sigma_{g} \times S^{1}} \frac{1}{2} \eps_{IJ}
\chi^{I}D_{t} \chi^{J} & \rightarrow &  \int_{\Sigma_{g} \times S^{1}}
\frac{1}{2} \eps_{IJ}(\f_{0})
\chi^{I}\, D_{t}(\f_{0}) \,  \chi^{J}  \\
\frac{1}{R}\int_{\Sigma_{g} \times S^{1}}\Omega_{IJKL} \,
\chi^{I}\chi^{J}\overline{\eta}^{K}  \eta^{L} &\rightarrow &
\int_{\Sigma_{g} \times S^{1}} \Omega_{IJKL} \, \chi^{I} \, \chi^{J}
\, \overline{\eta}^{K}_{0} \, \eta^{L}_{0} ,
\eea
and
\bea
& & \frac{1}{R}\int_{\Sigma_{g} \times S^{1}} \eps_{IJ}
\overline{\eta}^{I} D \chi^{J} \nonumber \\
& & \;\;\; \rightarrow  \int_{\Sigma_{g} \times
S^{1}} \left( \eps_{IJ}(\f_{0}) \overline{\eta}^{I}_{\perp} +
\overline{\eta}^{I}_{0 } \eps_{LJ}(\f_{0}) \f^{K}_{\perp}
\Gamma^{L}_{KI}(\f_{0}) \right)d \chi^{J} . \label{last}
\eea
So the combined actions (\ref{first}) to (\ref{last}) are what we end
up with. However, one can simplify matters greatly by noticing
that the path integrals over $\overline{\eta}^{I}_{\perp}$ and
$\eta^{I}_{\perp} $ imply that $\chi^{I}$ is harmonic on
$\Sigma_{g}$. We can feed this back in to the actions above to arrive
at rather more simplified expressions.

The action that one obtains on taking the limit is the sum of the
following two
\bea
S_{0} &= & \oint dt \left
( \frac{1}{2}g_{ij}(\f_{0})\partial_{t}\f^{i}_{0}\partial_{t}\f^{j}_{0} +
\eps_{IJ}(\f_{0}) \overline{\eta}^{I}_{0}D_{t}  \eta^{J}_{0}
\right. \nonumber \\
& & \left. \;\;\;\;\;\; -
\frac{1}{2} \eps_{IJ}(\f_{0})
\chi^{I}_{0}\, D_{t}(\f_{0}) \,  \chi^{J}_{0}  + \Omega_{IJKL} \,
\chi^{I}_{0}  \, \chi^{J}_{0}
\, \overline{\eta}^{K}_{0} \, \eta^{L}_{0}\right)
\eea
and
\bea
S_{\perp} &=& \int_{\Sigma_{g} \times S^{1}} \left[
\sqrt{h_{2}}\left(\frac{1}{2}g_{ij}(\f_{0}) 
\partial_{\mu}\f^{i}_{\perp}\partial^{\mu}\f^{j}_{\perp} + \eps_{IJ}(\f_{0})
\chi^{I}_{\perp \, \mu}D^{\mu}(\f_{0})  \eta^{J}_{\perp}\right)
\right. \nonumber \\
& & \;\;\;\;\; \;\;\;\;\;\;\; \;\;\;\;\;\;\;\;\; \left. + \; 
\eps_{IJ}(\f_{0}) \overline{\eta}^{I}_{\perp} D(\f_{0})
\chi^{J}_{\perp}\right] , 
\eea
Notice that $S_{0}$ is a standard, topological,
supersymmetric quantum mechanics action. 

After all these manouvers one obtains the partition function,
\be
Z_{X}^{RW}[\Sigma_{g}\times S^{1}] = \int D\f^{i}_{0}\,
D\eta^{I}_{0}\, D\overline{\eta}^{I}_{0} \, D\chi_{0}^{I}\, \ex{-\la S_{0}} \,
Z_{\perp}[\Sigma_{g}\times S^{1}] ,
\ee
where
\be
Z_{\perp} = \int D\f^{i}_{\perp} D\chi^{I}_{\perp} D\eta^{I}_{\perp}
D\overline{\eta}^{I}_{\perp} \, \, \ex{ - S_{\perp}} , \label{zperp}
\ee
and $\la$ is the volume of the $\Sigma_{g}$ with respect to the metric
$h_{2}$. The partition function (\ref{zperp}) is unity. As explained in
\cite{rw} this partition function calculates the Ray-Singer
torsion. Since we are integrating over modes which are not harmonic on
$\Sigma_{g}$ the cohomology that is being seen on $\Sigma_{g}\times S^{1}$ is
essentially acyclic. In this case the Ray-Singer torsion is honestly
`trivial' and the partition function is unity. 

We are left with the following partition function
\be
Z_{X}^{RW}[\Sigma_{g} \times S^{1}] = \int D\f^{i}_{0}\,
D\eta^{I}_{0}\, D\overline{\eta}^{I}_{0} \, D\chi_{0}^{I} \, \ex{- \la
S_{0}}  . \label{path}
\ee

\subsection{The Path Integral on $S^{2}\times S^{1}$}

On the two sphere there are no $\chi^{I}$ harmonic modes so that the
path integral becomes
\be
Z_{X}^{RW}[S^{2} \times S^{1}] = \int D\f^{i}_{0}\,
D\eta^{I}_{0}\, D\overline{\eta}^{I}_{0} \,  \ex{- \la
S_{0}'}  . 
\ee
where
\be
S_{0}' = \oint dt \left
( \frac{1}{2}g_{ij}(\f_{0})\partial_{t}\f^{i}_{0}\partial_{t}\f^{j}_{0} +
\eps_{IJ}(\f_{0}) \overline{\eta}^{I}_{0}D_{t}  \eta^{J}_{0}
\right) .
\ee

Happily enough, we do not have to evaluate this path
integral\footnote{However, it is not difficult to do so. Following
standard calculations as in \cite{ag} one obtains the second equality in
(\ref{dolb}) as it is given. One thing I made use of though, is the
isomorphism between the holomorphic and anti-holomorphic tangent
bundles. This is achieved by the covariantly constant, non-degenerate, tensor
$S^{\bar{I}}_{J} = g^{K \bar{I}}\eps_{KI}$. This means that one may
exchange $\eps_{IJ}\overline{\eta}^{J}$ for $g_{I\overline{J}}\, 
\overline{\eta}^{\overline{J}}$ in the supersymmetric quantum
mechanics theory.}. When $X$ is compact and hyper-K\"{a}hler 
this is the supersymmetric quantum mechanics path integral that
evaluates (minus\footnote{The sign is not easy to determine
apriori. However, with sign for the calculations on $T^{3}$ given, the
sign here follows from the calculations performed on Mapping-Tori})
the index of the Dolbeault operator \cite{ag,wsqm}. When
$X$ is hyper-K\"{a}hler but not compact, one finds the same
combination of Riemann tensor terms
as in the compact case (though this need no longer be the `index' of the
Dolbeault operator), hence\footnote{The definition of the Todd class is
\be
\prod_{i} \ex{(x_{i}/2)}\frac{(x_{i}/2)}{\sinh{(x_{i}/2)}} =
\ex{\sum_{i} (x_{i}/2)}\prod_{i} \frac{(x_{i}/2)}{\sinh{(x_{i}/2)}} =
\ex{c_{1}(X)/2} \prod_{i} \frac{(x_{i}/2)}{\sinh{(x_{i}/2)}} \nonumber
\ee
but for the manifolds in question one has $c_{1}(X)=0$. This means
that, $\hat{{\mathrm A}}(TX_{{\Bbb C}})= {\mathrm Todd}(TX_{{\Bbb
C}})$ a familiar fact for Calabi-Yau manifolds.}
\bea
Z_{X}^{RW}[S^{2}\times S^{1}] &=& -\int_{X} \, {\mathrm
Todd}\left(TX_{{\Bbb C}}\right) \nonumber \\
&=& -\int_{X} \, \prod_{i} \frac{(x_{i}/2)}{\sinh{(x_{i}/2)}}, \label{dolb}
\eea
where the $x_{i}$ are the eigenvalues of
\be
R_{I \overline{J}}= \frac{i}{2\pi} R_{I \overline{J}K \overline{L}}
\; dz^{K}d\overline{z}^{\overline{L}} \label{R}.
\ee

This calculation corroborates that in \cite{rw} which uses a Hilbert
space approach to determine the partition function. The advantage that
we have here is that we do not deal with the cohomology groups
directly and so do not have to identify $Z_{X}[S^{2}\times
S^{1}]$ with $\sum_{k=0}^{2n}(-1)^{k+1}{\mathrm dim}H^{(0,k)}$, though
this identification is correct for $X$ compact.

\subsection{The Path Integral on $T^{2}\times S^{1}$}

The sigma model in this case is rather easy to get a handle
on. Essentially one can forget all the non-harmonic modes so that the
final `path integral' is an integral over $X$ plus integration over
the fermions. The formula that one obtains is precisely that that one
would obtain from supersymmetric quantum mechanics for the index of
the de-Rham operator, that is the integral of the Euler class. 

In detail one sees that on $T^{2}$ there are two $\chi^{I}$ harmonic
modes, $\chi^{I}_{1}$ and $\chi^{I}_{2}$. One can combine fields in
the following way
\bea
\psi &= &\left( \chi_{1}^{I} , g^{\overline{I}
K}\eps_{KJ}\overline{\eta}^{J}_{0} \right) \nonumber \\
\overline{\psi} & = & \left( \eta^{I}_{0}, g^{\overline{I}
K}\eps_{KJ}\chi_{2}^{J} \right),
\eea
and the action in these variables is precisely that of the topological field
theory that calculates the Euler characteristic of $X$! The path integral
presents us with the Euler characteristic in Gauss-Bonnet form. Hence,
regardless of whether $X$ is compact or not, the final formula for the
path integral is as an integral over $X$ of the Euler class. Again this agrees
with the calculations in \cite{rw} for compact manifolds, namely that
\bea
Z_{X}^{RW}[T^{3}] &=& {\mathrm Str} \oplus_{i,j = 0}^{2n} H^{(i,j)}(X)
\nonumber \\
& = & \sum_{i,j = 0}^{2n} (-1)^{i+j} b^{(i,j)}(X) \nonumber \\
& = & \chi(X).
\eea

Now the Riemann curvature two-form $R^{a}_{\; b}$ is self dual for
both $K3$ and $X_{AH}$, indeed for all hyper-K\"{a}hler four
manifolds, so that the ratio of integrals that formally represent the
Euler characteristic and the $\hat{{\mathrm A}}$ genus do not depend on the
hyper-K\"{a}hler manifold in question. This is one of the main
properties exploited in \cite{rw} so as to use results from the
compact manifolds to arrive at equivalent statements for the
non-compact ones. Of course when the real dimension of $X$ is greater
than $4$, the situation becomes somewhat more involved.

\subsection{The $\chi$ Path Integral}

The number of components of $\chi_{0}$ depends on the genus $g$ of
$\Sigma_{g}$. For the sphere there are no such modes and the path
integral calculated the integral of the Todd class of $X$. This, for a
compact manifold, coincides with the index of the Dolbeault operator
$\overline{\partial}$ on $X$. So the $\eta$, $\overline{\eta}$ path
integral can be said to correspond to the $\overline{\partial}$
operator. On the Torus one found instead, owing to
the presence of two $\chi_{0}$ zero-modes, that the path integral
yields the integral over $X$ of the Euler class. This for compact $X$
corresponds to the index of the de-Rham operator. Alternatively this
may also be viewed as the index of the Dolbeault operator with values
in ${\mathrm T}^{(0,1)}X$. From this second, character valued,
viewpoint the $\eta$, $\overline{\eta}$ system still corresponds to
the Dolbeault operator, and the $\chi$ fields are there to take into
account the fact that the forms take values in ${\mathrm T}^{(0,1)}X$. 

For genus $g$ the $\chi$ system corresponds to forms
taking values in $\left(\wedge{\mathrm T}^{(0,1)}X\right)^{\otimes
g}$. Consequently, the path integral for compact $X$ calculates the
super dimension of 
\be
{\mathcal H}_{\Sigma_{g}} = \sum_{i}^{{\mathrm dim}_{{\Bbb C}}X}
H^{(i,0)}\left( X, \left(\wedge{\mathrm T}^{(0,1)}X
\right)^{ \otimes g} \right) , \label{homgp}
\ee
where the $H^{(i,0)}\left( X, \left(\wedge{\mathrm T}^{(0,1)}X
\right)^{ \otimes g} \right)$ are the Dolbeault cohomology groups of
$X$ with values in $\left(\wedge{\mathrm T}^{(0,1)}X\right)^{\otimes
g}$. The path integral gives the index of the operator in terms of
powers of the Riemann curvature tensor. When $X$ is non-compact
precisely the same combinations of the Riemann curvature tensor
appear. The only thing lacking is the interpretation of this object
as the index of the Dolbeault operator. 

In either case ($X$ compact or not) for $g > 1$ the path integral
vanishes. We saw this in the perturbative expansion and it is
unfortunate, but still true in the current setting since we can still
only pull down $R^{2n}$ (because of $\eta$ zero modes). It is unfortunate
for while one sees quite directly that the relevant Hilbert space (for
$X$ compact) is (\ref{homgp}) the usefulness, at the moment, of this remains
academic as the path integral vanishes. The fact that the path
integral vanishes is that the integral is over $X$ but the `form
degree' is greater than $4n$ (measured now by the number of $\eta$,
$\overline{\eta}$ modes and half of the $\chi$'s).

The details of the supersymmetric path integral approach to the index
theorem with values in a bundle (the $\chi$ integral) can be found,
for example, in \cite{ag}.

\section{Calculations on Mapping Tori}

In this section I will generalize slightly the three manifolds that
can be dealt with by shrinking a Riemann surface away. Here we will be
interested in Mapping Tori over a circle. A Mapping Torus is a three
manifold that is constructed from
$\Sigma_{g} \times [0,1]$ on gluing the two $\Sigma_{g}$ boundaries
together after acting on one of them by a diffeomorphism $f$.

The
path integral on such a manifold is the same as a path integral on
$\Sigma_{g} \times {\Bbb R}$ but with all fields $\Phi$ satisfying
\be
\Phi(x,t+1) = f^{*}\Phi(x,t) . \label{bc}
\ee
We still have the freedom, in the Rozansky-Witten model, to take the
zero volume limit of $\Sigma_{g}$ in the  path integral on $\Sigma_{g}
\times [0,1]$. This
will once more `squeeze' away all the states on $\Sigma_{g}$ except
for the harmonic modes.  Hence, the  path integral on $\Sigma_{g}
\times [0,1]$ with the boundary conditions (\ref{bc}) will devolve to
a path integral on the circle with the insertion of an operator that
implements the diffeomorphism on the harmonic modes. Denote this
operator by $U$ and denote the Mapping Torus by $\Sigma_{U}$.

What is $U$? We only need to ask how $f$ acts on the
$\chi_{0}^{I}$ fields since it
acts trivially on $\eta^{I}_{0}$ and $\overline{\eta}^{I}_{0}$. Let
$a_{ \a}$, and $b_{ \a}$, for $\a= 1,
\dots , g$, be a ``canonical'' basis for $H_{1}\left(
\Sigma_{g} , {\Bbb Z} \right)\cong {\Bbb Z}^{2g}$, such that the
intersection pairing of the cycles satisfies
\be
(a_{ \a}, a_{ \b}) = (b_{ \a}, b_{ \b}) =0 ,
\ee
and
\be
(a_{ \a}, b_{ \b}) = -(b_{ \a}, a_{ \b}) = \d_{\a \b}.
\ee
Let $\o^{\a}_{a}$ and $\o^{\a}_{b}$ be a basic set of
real harmonic $1$-forms dual to the homology basis
\be
\int_{a_{ \b}} \, \o^{\a}_{a} = \int_{b_{ \b}} \, \o^{\a}_{b} =
\d^{\a \b} , \;\;\; 
\int_{a_{ \b}} \, \o^{\a}_{b} = \int_{b_{ \b}} \, \o^{\a}_{a}
=0,\label{norm}
\ee
and
\be
\int_{\Sigma}\o^{\a}_{a} \wedge \o^{\b}_{b} = \sum_{\gg =1}^{g} 
\left( \int_{a_{\gg}}  \o^{\a}_{a} \int_{b_{\gg}} \o^{\b}_{b} -
\int_{a_{\gg}} \o^{\b}_{b}\int_{b_{\gg}} \o^{\a}_{a} \right) = \d^{\a
\b}. 
\ee
The mapping class group acts on the homology basis by elements of
${\mathrm Sp}\left(2g , {\Bbb Z}\right)$. Let
\be
E = \left(
\begin{array}{cc}
0 & {\Bbb I} \\
-{\Bbb I} & 0
\end{array} \right)
\ee
a $2g \times 2g $ matrix. Then $U \in {\mathrm Sp}\left(2g , {\Bbb
Z}\right) $ means that
\be
U^{{\mathrm T}}\, .E.\, U = E ,
\ee
and one has
\be
U = \left( \begin{array}{cc}
A & B \\
C & D
\end{array}\right) ,
\ee
where
\bea
A^{{\mathrm T}}C - C^{{\mathrm T}}A &=& 0 \nonumber \\
B^{{\mathrm T}}D - D^{{\mathrm T}}B &=& 0 \nonumber \\
A^{{\mathrm T}}D - C^{{\mathrm T}}B & = & {\Bbb I}.
\eea

Now expand $\chi^{I}_{0}$ in the basis,
\be
\chi^{I}_{0} = \chi^{I \, a}_{\a}(t)\o^{\a}_{a} + \chi^{I\, b}_{\a}(t)
\o^{\a}_{b} ,
\ee
where the $\chi^{I}_{\a}$ are fields on the $S^{1}$.  By virtue of (\ref{norm})
the action of the mapping class group, thought of as acting on the
$\chi^{I}$ is exactly the same as on the homology basis. For $U \in
{\mathrm Sp}\left(2g , {\Bbb Z}\right) $ the action on $\chi^{I}$
is
\be
\left( \begin{array}{c}
\chi^{I}_{a} \\
\chi^{I}_{b} 
\end{array} \right) \rightarrow 
\left( \begin{array}{c}
A\chi^{I}_{a} + B \chi^{I}_{b}\\
C\chi^{I}_{a} + D \chi^{I}_{b}
\end{array} \right).
\ee

\subsection{The Path Integral}
The path integral to be performed,
\be
Z^{RW}_{X}[\Sigma_{U}] = \int D\f^{i}\,
D\eta^{I}\, D\overline{\eta}^{I} \, \int_{U}D\chi^{I} \, \ex{- \la
S_{0}}  , \label{tpath}
\ee
has the same form as (\ref{path}) except that the $\chi^{I}$ are
now not periodic on the circle but rather satisfy twisted boundary
conditions. The zero subscript on the fields has been dropped as it is
clear that
this path integral is on the circle and all tensors on $X$ are
understood to depend on $\f_{0}$. The path integral measure is
\be
\left(D\f^{i}\sqrt{\la}\right)\left(D\eta^{I}\sqrt{\la}\right)
\left(D\overline{\eta}^{I} \sqrt{\la}\right) \left(D\chi^{I}
\sqrt{\la}\right) . 
\ee
We expand the fields in Fourier modes on the circle and perform that
following scalings; for $\f^{i}$, $\eta^{I}$ and $\overline{\eta}^{I}$
all their modes except the constant mode are scaled (divided) by
$\sqrt{\la}$, all the modes of $\chi^{I}$ are scaled by
$\sqrt{\la}$. The measure is now $\la$ independent and the path
integral factorises as
\be
Z^{RW}_{X}[\Sigma_{U}] = \int D\f^{i}\,
D\eta^{I}\, D\overline{\eta}^{I} \ex{-S'} \; \int_{U} D\chi^{I}
\ex{-S_{\chi}} , \label{fact}
\ee
where
\bea
S_{\chi} & = &  \oint \left(
\frac{1}{2} \eps_{IJ}
\chi^{I}\, \partial_{t} \,  \chi^{J}  + \Omega_{IJKL} \,
\chi^{I} \, \chi^{J}
\, \overline{\eta}^{K}_{0} \, \eta^{L}_{0} \right) \nonumber \\
& = & \oint \left(
\frac{1}{2} \eps_{IJ} \, 
(\chi^{{\mathrm T}})^{I}. E . \, \partial_{t} \,  \chi^{J}  +  
\Omega_{IJKL} \,
(\chi^{{\mathrm T}})^{I} . E .  \chi^{J}
\, \overline{\eta}^{K}_{0} \, \eta^{L}_{0} \right)\label{chiact}
\ee
and the $\eta^{I}_{0}$, $\overline{\eta}^{K}_{0}$ are the
constant modes of the fields. 

It is not completely straightforward to evaluate the $\chi$ path
integral. The procedure that I will follow is to change variables from
the $\chi^{I}$ fields to a periodic set of fields
$\psi^{I}$. The price to be paid is that a new connection appears in
the action for the $\psi^{I}$ fields. Let
\be
\chi^{I}(t) = \ex{itv}.\psi^{I}(t) , \label{newpsi}
\ee
where
\be
\ex{iv} = U .
\ee
As one varies $t$, the matrix $\ex{it v}$ runs along a path in $Sp(2g, {\Bbb
R})$ from the identity to $U \in Sp(2g, {\Bbb Z})$. The field
$\psi^{I}$ is periodic in $t$, so that
\bea
\chi^{I}(t+n) &= & \ex{i(t+n)v}.\psi^{I}(t) \nonumber \\
&= & U .\chi^{I}(t+n-1) ,
\eea
as required on the Mapping Torus. The action (\ref{chiact}), in terms
of the new variables is,
\be
S_{\psi} =  \oint \, (\psi^{{\mathrm T}})^{I} . E. \left(
\frac{1}{2} \eps_{IJ} \, 
\, (\partial_{t} \,  + i v ) + \Omega_{IJKL} \, \overline{\eta}^{K}_{0} \,
\eta^{L}_{0} \right) \psi^{J} , \label{psiact}
\ee
so that the path integral goes over to
\be
\int_{U} D\chi^{I}
\ex{-S_{\chi}} = \int D\psi^{I}\ex{-S_{\psi}} \label{psipi}.
\ee
Formally this path integral evaluates the square root of the
determinant (Pfaffian) of the operator
\be
(\partial_{t} \,  + i v )\d^{I}_{J} + R^{I}_{JKL} \, \overline
{ \eta}^{K}_{0} \, \eta^{L}_{0} ,
\ee
which is the (pull-back of the) covariant derivative on sections of an
${\mathrm Sp}(2g, {\Bbb R}) \otimes {\mathrm Sp}(n)$ bundle. One may
write the operator then as
\be
\partial_{t} + A
\ee
where
\be
A = v \otimes {\Bbb I} \oplus {\Bbb I} \otimes R . \label{gaugef}
\ee
In (\ref{gaugef}) $R$ is the matrix form, 
\be
R_{I\overline{J}}(\eta, \overline{\eta}) =
\frac{i}{2\pi}R_{I\overline{J} K\overline{L}}
\eta^{K} \overline{\eta}^{\overline{L}} \label{Reta}.
\ee
Notice that the ${\mathrm Sp}(n)$ part of the gauge field is precisely
the component of (\ref{conn}) in the time direction. 

To proceed I will evaluate the determinant and then take its square
root. One way to do this is to work on the interval and to specify boundary
data. This approach is explained in a related context\footnote{At one
point in the derivation in \cite{btg/g} the fact that for a group
valued field $\det{({\mathrm Ad}g})_{{\bf k}}=1$ is used. The equivalent,
$\det(e^{R}) = 1$, in the present context holds as $c_{1}(X) =
0$, this being another point of contact with Chern-Simons theory.} in
section (3.1) of \cite{btg/g}.

By following the derivation in \cite{btg/g} one obtains
\be
\Det{\left(U \otimes {\Bbb I} - {\Bbb I} \otimes \ex{R(\eta ,
\overline{\eta})} \right) }^{1/2} \label{det}
\ee
which agrees with (the equivalent of) (3.24) of \cite{btg/g} when
$U=I$. 
The path integral over the non-constant modes of $\f^{i}$, $\eta^{I}$
and $\overline{\eta}^{I}$ give back the integrand of the $S^{2} \times
S^{1}$ path integral (\ref{dolb}) but with the $x_{i}$ now the
eigenvalues of (\ref{Reta}). The integral over the constant modes of
$\eta^{I}$ and $\overline{\eta}^{\overline{I}}$ turn the $x_{i}$ into
the eigenvalues of (\ref{R}). So, putting all the pieces together we
obtain
\be
Z^{RW}_{X}[\Sigma_{U}] = -\int_{X} \, {\mathrm
Todd}\left(TX_{{\Bbb C}}\right) \, \Det{\left(U\otimes I - I \otimes \ex{R}
\right) }^{1/2} . \label{mtf}
\ee
One can now write this as
\be
Z^{RW}_{X}[\Sigma_{U}] = c_{X} I[\Sigma_{U}].
\ee
To obtain the coefficient, $I[\Sigma_{U}]$, of the top form one essentially
differentiates the integrand, as a function of a variable $u$ ($R
\rightarrow u$), $2n$ times and evaluates at $u=0$. That coefficient is
the data that depends on the Mapping Torus while the integral over $X$
of the top form yields $c_{X}$. However, one can show that
$I[\Sigma_{U}]$ obtained in this way is related to the Alexander
polynomial of $I[\Sigma_{U}]$ (and its derivatives). This is a
special example of the more general result that for all three
manifolds with $b_{1}=1$, the partition function is a function of the
Alexander polynomial of that manifold. This last fact is established
in \cite{ht} using a path integral argument that is quite different to
the one employed here. 

There are some easy checks that one can make on (\ref{mtf}). Firstly
when $\Sigma_{g}=S^{2}$ the determinant is formally unity. This gives
us back (\ref{dolb}). When
$\Sigma_{g} = T^{2}$, with $U = {\Bbb I}$, the determinant is
\bea
\Det{\left( {\Bbb I}\otimes {\Bbb I} - {\Bbb I}\otimes
\ex{R}\right)^{1/2}} & = & \Det{\left( {\Bbb I}- \ex{R} \right)}
\nonumber \\
& = & -\prod 2\sinh{\left( x_{i}/2\right)} ,
\eea
which means that
\be
-{\mathrm Todd}\left(TX_{{\Bbb C}}\right) \Det{\left( {\Bbb I}\otimes
{\Bbb I} - {\Bbb I}\otimes \ex{R}\right)^{1/2}}= \prod x_{i} =
{\mathrm e}(X).
\ee
This reproduces the the result for $T^{2}$. Notice, however, that
there is an ambiguity in the choice of the root of the
determinant. The choice I have made is consistent with the formula,
derived below, for the mapping torus. There the sign is set by
demanding that it agrees with the formula for $S^{2}\times S^{1}$.

Recall that for a curve of
genus greater than one the quantum
mechanics path integral for $\Sigma_{g}\times S^{1}$ vanishes as one
is obtaining too high powers of the curvature two form. In the present
situation, we can see the same result for the same reason. From the
general formula (\ref{mtf}) one
finds that with $U= {\Bbb I}$,
\be
\Det{\left( {\Bbb I}\otimes {\Bbb I} - {\Bbb I}\otimes
\ex{R}\right)^{1/2}}  =  \Det{\left( {\Bbb I}- \ex{R} \right)^{g}}
\ee
the determinant starts off as $\left(\prod x_{i}\right)^{g}$, which
vanishes when $g \geq 2$.

\subsection{The Casson Invariant for Mapping-Tori}

When ${\mathrm dim}_{{\Bbb R}}X=4$, we are (essentially) calculating
the Casson invariant. To begin with let the Riemann surface be a
Torus. In such a situation the
Mapping-Torus is a Torus bundle over the circle. The matrix U, is now
an element of $SL(2, {\Bbb Z})$ and takes the form
\be
U = \left(
\begin{array}{cc}
p & q \\
r & s 
\end{array} \right), \;\; {\mathrm with}\; \;  ps-qr = 1,\; {\mathrm and}
\;\;\;  p,q,r,s \in {\Bbb Z}.
\ee
One calculates that
\bea
\Det{\left(U\otimes {\Bbb I} - {\Bbb I} \otimes \ex{R}
\right) }^{1/2} & =&  \Det{\left(
\begin{array}{cc}
p{\Bbb I}- \ex{R} & q{\Bbb I} \\
r {\Bbb I}& s {\Bbb I} -\ex{R}
\end{array} \right)}^{1/2} \nonumber \\
& = & \Det{\left({\Bbb I} - (p+s)\ex{R} + \ex{2R} \right)}^{1/2}
\nonumber \\ 
& = & \Det{\left(2\cosh{R} - (p+s){\Bbb I}\right)}^{1/2}. \label{mtdets}
\eea

When ${\mathrm dim}_{{\Bbb R}}X=4$ the (2-form) eigenvalues of $R$ are $x$
and $-x$ (since the manifold is Ricci flat). This means that in this
case we have
\be
\Det{\left(2\cosh{R} - (p+s){\Bbb I}\right)}^{1/2} = \left(2\cosh{x} -
(p+s)\right) \, ,
\ee
and
\bea
Z_{X}^{RW}[T^{2}_{U}] &=& \int_{X} -\frac{x^{2}}{4\sinh^{2}{x/2}}
\left(2 \cosh{x} - (p+s)\right) \nonumber \\
&=& - \frac{1}{2}\int_{X} x^{2} \, . \left( \frac{u^{2}}{4\sinh^{2}{u/2}}
(2 \cosh{u} - (p+s)) \right)_{u=0}'' \nonumber \\
& = & c_{X} \, I[T^{2}_{U}],
\eea
where
\bea
I[T^{2}_{U}] &=& -\left( \frac{u^{2}}{4\sinh^{2}{u/2}}
(2 \cosh{u} - (p+s)) \right)_{u=0}''  \nonumber \\
& = & -\frac{1}{6}(p + s + 10),
\eea
and
\be
c_{X} = -\frac{1}{4} b_{\theta}(X) ,
\ee
with
\be
b_{\theta}(X) = -2\int_{X}  x^{2} .
\ee
All of this is in complete agreement with the calculations of Rozansky
and Witten who arrive at these formula in two different (from the
present derivation) ways. Their first method is to use the result from
Chern-Simons theory that one is calculating the second derivative of
the Ray-Singer torsion of the manifold while the second method is to
make use of the explicit action of $U$ on the Hilbert space of states
when $X$ is a K$3$ surface.

For hyper-K\"{a}hler manifolds of real dimension $4$ and
for $\Sigma_{g}$ with any genus $g$,
\bea
Z_{X}^{RW}[\Sigma_{U}] &=&  - \int_{X} {\mathrm Todd}\left(TX_{{\Bbb
C}} \right){\mathrm Det}\left( U-{\Bbb I}\right)-\int_{X} \Det{\left(U\otimes {\Bbb I} -
{\Bbb I} \otimes \ex{R} \right) }^{1/2}  \nonumber \\
&=&  c_{X} \left( \frac{1}{6} + {\mathrm Tr}(U-{\Bbb I})^{-1} +
{\mathrm Tr}(U-{\Bbb I})^{-2} \right) {\mathrm Det}\left( U-{\Bbb I}\right).
\eea

\section{Reduction of the Rozansky-Witten Theory to Two Dimensions}\label{red}

Given a theory on a $d+n$ dimensional manifold $Y_{d+n}$, there are
two different ways to obtain a theory on a $d$ dimensional manifold
$M_{d}$. The
first is to consider a manifold of the form $Y_{d+n}=M_{d}\times
{\Bbb R}^{n}$ and to simply to `forget' the dependence of the fields on the
${\Bbb R}^{n}$ coordinate-this goes by the name of dimensional
reduction. The second is to consider the theory on a manifold
$Y_{d+1}=M_{d} \times N_{n}$ and then to perform a harmonic eigenmode 
expansion of the fields with respect to some suitable operator on
$N_{n}$ say, for example, in terms of eigenmodes of the Laplacian. The
second approach is called Kaluza-Klein reduction, and yields a theory
with a finite number of fields on $M_{d}$, when it is possible to
integrate out most of the infinite tower of fields (one for each
eigenmode). Typically, in conventional field theory, one takes the
size of $N_{n}$ to be very small and then almost all the modes decouple. The
error in doing this goes like the size of $N_{n}$, which for
physical reasons (we do not see it) is very small. In a
topological field theory one may have the freedom to vary the volume
at will (since the theory should be metric independent) and therefore
pass from the theory in $d+n$ dimensions to the theory
in $d$, without error. 

For the  Rozansky-Witten theory on a three manifold of the form
$\Sigma_{g} \times S^{1}$ there are then a number of approaches that
one may take. One can, as in section \ref{pert}, evaluate the path
integral perturbatively by using the freedom to scale the coupling
constant in front of
the action which leads to the expressions already determined in terms of
Greens functions on the three manifold. We have also used the
Kaluza-Klein idea in order to equate the path integral of the
Rozansky-Witten theory on
$\Sigma_{g}\times S^{1}$ with a supersymmetric field theory on
$S^{1}$. Alternatively one can, instead,
shrink the radius of the $S^{1}$, that is one may perform a
Kaluza-Klein reduction on the circle, to obtain an effective theory on
$\Sigma_{g}$ that should also be equivalent to the Rozansky-Witten theory.

What theory will be obtained when one dimensionally reduces? There are two
topological sigma models known in two dimensions that arise from the
twist of the
standard supersymmetric sigma model. They can be reduced to
integrals over the moduli space of pseudo-holomorphic curves in the case
of the ${\mathbf A}$-model or to an integral over constant maps in the
case of the ${\mathbf B}$-model. A moments
reflection will show that it must be the topological sigma model known
as the ${\mathbf B}$-model \cite{wms} that we obtain on reduction of
the Rozansky-Witten theory, since there too the path integral is
reduced to an integral over constant maps. We can be more systematic
about the relationship.

First the field content is the same. In a given complex structure of
the hyper-K\"{a}hler manifold $X$, let $\f^{I}$ be local holomorphic
coordinates. The fields appearing in the ${\mathbf B}$-model are
bosonic maps, $\f^{i}$ from $\Sigma$ to $X$. There is also a Grassmann
odd one form $\rho$ with values $\f^{*}\left({\mathrm T}^{(1,0)}X
\right)$ and two Grassmann odd zero forms with values in
$\f^{*}\left({\mathrm T}^{(0,1)}X \right) $. This is precisely the content
of the Rozansky-Witten model on dimensional reduction. The bosonic
field $\f^{i}$ is clearly there. In the prefered complex structure one
has $V= \f^{*} \left({\mathrm T}^{(1,0)}X \right)$, so that $\eta$ is
one of the
Grassmann odd sections of $\f^{*}\left({\mathrm T}^{(0,1)}X \right) $ while
the component of $\chi$ in the $S^{1}$ direction provides the
other. The one form is supplied by the rest of $\chi$. To make precise
contact with the fields in \cite{wms} one makes use of the natural
isomorphism between the holomorphic and anti-holomorphic tangent
bundles of the hyper K\"{a}hler manifold $X$. One sets 
\be
\eta^{\overline{I}} = g^{\overline{I}J}\eps_{JK}\eta^{K}, \;\;\;
\theta_{I} = \eps_{IJ}\overline{\eta}^{J}, \;\;\; \rho^{I} =
\chi^{I}. \label{ids}
\ee
Secondly one can compare the supersymmetry transformations,
(\ref{brst}), after reduction directly with $(4.2)$ in \cite{wms} and
see, with the identifications made in (\ref{ids}), that they agree.

Of course, the ultimate test is that the actions agree as
indeed they do. We can read off from (\ref{sig1},\ref{sig2}) the
dimensionally reduced actions,
\bea
S_{1} & = & \int_{\Sigma_{g}} \, \sqrt{h_{2}} \,
\frac{1}{2} g_{ij}\partial_{\mu}\f^{i}\partial^{\mu}\f^{j} +
 \eps_{IJ} \chi^{I}_{\mu}D^{\mu} \eta^{J}  \label{siga}\\
S_{2} & = & \int_{\Sigma_{g}} \, \left(\eps_{IJ}
 \overline{\eta}^{I} D \chi^{J} 
+ \Omega_{IJKL} \chi^{I}\chi^{J}\overline{\eta}^{K} \eta^{L} \right) ,
\label{sigb}
\eea
which are in agreement with $(4.3)$, $(4.4)$ and $(4.5)$ of
\cite{wms}.

There is, in principle, an anomaly in the ${\mathbf B}$-model as the
fields that make up the quadratic part of the Grassmann action are
sections of different bundles. The fermionic determinant will thus
have an anomaly unless certain conditions are met. Indeed it turns out
that $X$ must satisfy $c_{1}(X)=0$. The fact that the manifold $X$
that appears in the Rozansky-Witten theory is hyper-K\"{a}hler ensures that the
${\mathbf B}$-model makes sense. 

We have not shown the equivalence of the generalized Casson invariant
on a three manifold $M= \Sigma \times S^{1}$ with the ${\mathbf
B}$-model, for the same target space $X$. To do that one would have to
show that shrinking the $S^{1}$ reproduces the ${\mathbf
B}$-model. They are, infact, not equivalent; see footnote \ref{foot}
for a brief discussion of this point.

\subsection{Observables}
As shown in \cite{wms} the observables of the ${\mathbf B}$-model are
naturally equivalent to,
\be
\oplus_{p,q} H^{(0,p)}\left(X, \bigwedge^{q} {\mathrm
T}^{(1,0)}X \right). 
\ee

As discussed by Witten \cite{wms} in order to describe the mirror map
from the point of view of the topological sigma models one needs to
thicken the moduli space a bit. One partial thickening is to define a
${\mathbf B}$ model which makes sense as one changes the complex
structure of $X$. This is what Witten called a classical
deformation. There are other thickenings that one may consider and one
such example may be found in \cite{wms}. Fortunately, for us, the
model of Rozansky and Witten comes equipped with a supersymmetry that
does not require the specification of the complex structure of
$X$. This property passes down to the ${\mathbf B}$-model of this
section. Hence, for a hyper-K\"{a}hler target space the ${\mathbf B}$
model (\ref{siga}, \ref{sigb}) is well defined for all the complex
structures that are compatible with the hyper-K\"{a}hler
structure. The essence of the matter is that the action enjoys two
linearly independent BRST symmetries the second BRST supersymmetry being,
\be
\begin{array}{ll}
Q\f^{I} = \eta^{I}, & Q\f^{\overline{I}} = 0, \\
Q\eta^{I}= 0, & Q\chi^{I} =
\eps^{IJ}g_{J\overline{K}}d\f^{\overline{K}} -
\Gamma^{I}_{JK}\eta^{J}\chi^{K} ,
\end{array} \label{brst2}
\ee
and, furthermore, it is exact with respect to both $Q$ and
$\overline{Q}$. $A_{\overline{I}}^{\;\; J}$, the element in $H^{(0,1)}(X,
T^{(1,0)}X)$ that deforms the complex structure while preserving the
hyper-K\"{a}hler structure, is proportional to $g_{\overline{I}
K}\eps^{KJ}$. 

One can now compare with the proposal of Labastida and Marino
\cite{lm} for perturbing the ${\mathbf B}$-model so that it
incorporates complex structure deformations. From the discusion of the
previous paragraph we should find that the perturbed ${\mathbf
B}$-model agrees with the unperturbed ${\mathbf B}$-model. They
proposed a deformed action together with a deformed BRST supersymmetry. That
supersymmetry, their equation (26), is a linear combination of the
$Q$ and $\overline{Q}$ supersymmetries and the action\footnote{To show
the equivalence of the transformation rules and of the action one
needs to remember that $D_{L}\left( g_{\overline{I} K}\eps^{KJ}\right)
=0$.}, their equation (32), agrees with (\ref{siga}, \ref{sigb}).

\subsection{Reduction of the Kapranov-Kontsevich Theory to Two
Dimensions} 

There is a refinement of the Rozansky-Witten theory due to Kontsevich
\cite{kon} and Kapranov \cite{kap}. The point is that one may lift the
requirement that $X$ be hyper-K\"{a}hler. Instead one considers $X$
to be a complex manifold with a holomorphic symplectic structure. This
means that $X$ comes equipped with a two form $\eps$ which is closed,
$d\eps = 0$, holomorphic (that is a (2,0) form) and of maximal rank
(non-degenerate), $\eps^{n} \neq 0$. Manifolds of this type also have
vanishing $c_{1}(X)$. The corresponding three dimensional
field theory, which requires a choice of Hermitian metric, is
described in the appendix of \cite{rw}. 

The model that one obtains is actually a generalisation of the
${\mathbf B}$-model as holomorphic symplectic manifolds do not have
to be K\"{a}hler and so, in particular, do not have to be Calabi-Yau
manifolds. Recall that the topological models are obtained by twisting
the $N=2$ sigma model. The extended supersymmetric theory is
formulated on a K\"{a}hler manifold and so by construction the
${\mathbf A}$ and ${\mathbf B}$ models are defined on K\"{a}hler
manifolds. However, the ${\mathbf A}$-model makes sense without the
K\"{a}hler condition on $X$, it is enough that $X$ admit an almost
complex structure. Consistency for the ${\mathbf B}$ model requires
that $c_{1}(X)=0$.

The action for the ${\mathbf B}$-model on a Riemann surface $\Sigma_{g}$, is
found just by dimensional reduction\footnote{Since the topological
theory should
not depend on the metric one puts on the three manifold one could also
consider the theory in three dimensions and take the radius of the
$S^{1}$ to zero. This will not give back the two dimensional
model. Rather one will obtain the two dimensional theory together with
the insertion of an operator. It is the expectation value of this
operator that will give back the Casson invariant. \label{foot}} of the model
presented in \cite{rw} and setting
$\chi_{t}^{I} = \overline{\eta}^{I}$ where $t$ is the $S^{1}$ co-ordinate,
\be
S_{{\mathbf B}} = c_{1}\int_{\Sigma} L_{1} +
c_{2}\int_{\Sigma}\, L_{2} , \label{b2}
\ee
where
\be
L_{1} = \{ Q , g_{I\overline{J}}\chi^{I}_{\mu}
\partial^{\mu}\overline{\f}^{\overline{I}} \}
\ee
and
\be
L_{2} &=& \eps_{IJ}\chi^{I}D \overline{\eta}^{J} - \frac{1}{3}
\eps_{IJ}R^{J}_{\,
KL\overline{M}}\chi^{I}\chi^{K}\overline{\eta}^{L}\eta^{\overline{M}}
- \frac{1}{6}
\eps_{IJ}R^{J}_{\,
KL\overline{M}}\overline{\eta}^{I}\chi^{K}\chi^{L}\eta^{\overline{M}}\nonumber \\
& & \;\;\;  +
\frac{1}{6}(\nabla_{L}\eps_{IK})d\f^{I}(\chi^{K}\overline{\eta}^{L} +
\chi^{L}\overline{\eta}^{K}) . \label{kk}
\eea
The BRST operator $Q$ acts in the following way
\be
\begin{array}{lll}
Q \f^{I}= 0, & Q \f^{\overline{I}} = \eta^{\overline{I}} &
Q\overline{\eta}^{I}=0 \\
Q \eta^{\overline{I}} = 0, & Q\chi^{I} = - d \f^{I} . & 
\end{array}
\ee

It was also pointed out in \cite{rw} that the BRST class of $L_{2}$ is
independent of the connection $\Gamma_{JK}^{I}$, that is, one can add
to $\Gamma_{JK}^{I}$ a tensor $A^{I}_{JK}$, and $L_{2}$ is then
changed by a BRST exact term. One immediate implication of this is
that it is BRST equivalent to work with a covariantly constant
$\eps$. We have to show that even though
\be
\nabla_{K}\left( \Gamma\right)\eps_{IJ} = \nabla_{K}\eps_{IJ}\neq 0,
\ee
one can choose a new connection $\nabla_{K}\left( \Gamma + A\right)$
so that
\be
\nabla_{K}\left( \Gamma + A \right)\eps_{IJ}= \nabla_{K}'\eps_{IJ}=0 .
\ee
Let $\hat{\eps}^{IJ}$ be the matrix that inverts $\eps_{IJ}$,
\be
\hat{\eps}^{IJ} \eps_{JK} = \d^{I}_{K} .
\ee
Such a matrix exits as $\eps$ is non-degenerate, furthermore it is
holomorphic since $\eps$ is. The required tensor $A^{I}_{JK}$ is
\be
A^{I}_{JK} = \frac{1}{3} \hat{\eps}^{IM}\left( \nabla_{J}\eps_{MK} +
\nabla_{K}\eps_{MJ} \right).
\ee
One finds,
\bea
\nabla_{K}'\eps_{IJ} &=& \nabla_{K}\eps_{IJ} - A^{M}_{KI}\eps_{MJ} -
A^{M}_{KJ}\eps_{IM} \nonumber \\
& = & \nabla_{K}\eps_{IJ} + \frac{1}{3}\left( \nabla_{K}\eps_{JI}+
\nabla_{I} \eps_{JK} \right) - \frac{1}{3}\left( \nabla_{K}\eps_{IJ}+
\nabla_{J}\eps_{IK} \right) \nonumber \\
& = & \frac{1}{3}\left( \nabla_{K}\eps_{IJ}+ \nabla_{I} \eps_{JK} +
\nabla_{J} \eps_{KI} \right) \nonumber \\
&=& \frac{1}{3}\left( \partial_{K}\eps_{IJ}+ \partial_{I}\eps_{JK} +
\partial_{J}\eps_{KI} \right) \nonumber \\
&=& 0,
\eea
and the last equality follows since $\eps$ is also
closed. 

Consequently, we see that there is precisely enough information
in having a holomorphic symplectic structure to be able to find a
connection with respect to which $\eps$ is covariantly constant. BRST
invariance tells us that we may work with such a connection without
changing the results of the topological field theory. As far as the
theory is concerned this means that we can just as well drop the last
line of (\ref{kk}), providing that we understand all covariant
derivatives to be with respect to the new connection. Once one has
chosen the connection so that the holomorphic symplectic
structure $\eps$ is also parallel then all the Chern
forms\footnote{Meaning that it is not only the (2j+1)th Chern numbers
that vanish, but the Chern-Weil representatives themselves.} of
$c_{2j+1}(X)$ vanish \cite{kob}. Some stringent
conditions on the cohomology of $X$ can be deduced from these facts,
\cite{kob}. 

Returning to the properties of the field theory, the scaling arguments
used to deduce that the Rozansky-Witten model devolves to an integral
over constant maps can be directly taken over to the
Kapranov-Kontsevich theory. The new ${\mathbf B}$-model (\ref{b2}),
likewise can be shown to devolve to an integral on the space of
constant maps, that is to an integral over $X$.

This generalised ${\mathbf B}$-model is probably obtained from the
twisting of an $N=2$ sigma model with non-zero $B$ field. In
particular, one expects that $B = \eps$.

\subsubsection*{Acknowledgements}
I would like to thank N. Habegger and M. Narasimhan and especially
M. Blau for discussions on this work.

\rnc{\Large}{\normalsize}

\end{document}